\documentclass[conference,a4paper]{APSIPA2020}
\usepackage{multirow}
\usepackage[dvips]{graphicx}
\usepackage{amsmath}
\usepackage[psamsfonts]{amssymb}
\usepackage{amsxtra}
\usepackage{threeparttable}

\renewcommand{\baselinestretch}{0.907}

\begin{document}
\title{Simultaneous measurement of time-invariant linear and nonlinear, and random and extra responses using frequency domain variant of velvet noise
}

\author{%
\authorblockN{%
Hideki Kawahara\authorrefmark{1},
Ken-Ichi Sakakibara\authorrefmark{2},
Mitsunori Mizumachi\authorrefmark{3},
Masanori Morise\authorrefmark{4} and
Hideki Banno\authorrefmark{5}
}
\authorblockA{%
~\hspace{2cm}\authorrefmark{1}
Wakayama University, Wakayama, Japan\ \ 
E-mail: kawahara@wakayama-u.ac.jp}
\authorblockA{%
~\hspace{-0.15cm}\authorrefmark{2}
Health Science University of Hokkaido, Sapporo, Japan\ \ 
E-mail: kis@hoku-iryo-u.ac.jp}
\authorblockA{%
~\hspace{0.54cm}\authorrefmark{3}
Kyushu Institute of Technology, Kitakyushu, Japan\ \ 
E-mail: mizumach@ecs.kyutech.ac.jp}
\authorblockA{%
~\hspace{3.46cm}\authorrefmark{4}
Meiji University, Tokyo, Japan\ \ 
E-mail: mmorise@meiji.ac.jp}
\authorblockA{%
~\hspace{3.12cm}\authorrefmark{5}
Meijo University, Nagoya, Japan\ \
E-mail: banno@meijo-u.ac.jp}
}

\maketitle
\thispagestyle{empty}

\begin{abstract}
We introduce a new acoustic measurement method that can measure the linear time-invariant response, the nonlinear time-invariant response, and random and time-varying responses simultaneously. The method uses a set of orthogonal sequences made from a set of unit FVNs (Frequency domain variant of Velvet Noise), a new member of the TSP (Time Stretched Pulse). FVN has a unique feature that other TSP members do not. It is a high degree of design freedom that makes the proposed method possible without introducing extra equipment. We introduce two useful cases using two and four orthogonal sequences and illustrates their use using simulations and acoustic measurement examples. We developed an interactive and realtime acoustic analysis tool based on the proposed method. We made it available in an open-source repository. The proposed response analysis method is general and applies to other fields, such as auditory-feedback research and assessment of sound recording and coding.
\end{abstract}

\section{Introduction}
We introduce a measurement procedure that simultaneously analyzes linear, nonlinear, and random attributes of the system.
It uses a new member of the TSP (Time Stretched Pulse), called FVN (Frequency domain variant of Velvet Noise)\cite{kawahara2018IS}.
We extended our previous tools for measuring system responses using FVN\cite{kawahara2019apsipa}.
The extension takes advantage of the unique feature of FVN, a high degree of design parameters that provides arbitrarily many orthogonal sequences.
These orthogonal sequences make it possible to measure the linear time-invariant response, 
the nonlinear time-invariant response, and random and temporally variable responses simultaneously.
We introduce applications of this new formulation for acoustic measurement of loudspeaker systems and demonstrated using simulations and actual acoustic measurement examples.
The proposed method is general and applies to other fields that need analysis of systems consisting of linear, nonlinear, and random components.

This article has the following organization.
It starts by reviewing TSP-based measurements and issues in acoustic measurements.
In section~\ref{ss:unitFVN}, we introduce the unit FVN and its application to the impulse response measurement.
In section~\ref{ss:twoFVNseq}, we introduce orthogonal sequence generation from unit FVNs and orthogonal binary sequences.
We focus on two cases, two sequences, and four sequences because they are practically useful.
Using two orthogonal sequences provides simultaneous measurement of two acoustic paths and introduces the essential idea of orthogonalization.
Section~\ref{ss:fvnMes4seq} generalizes the idea for four sequences.
Using four orthogonal sequences provides simultaneous measurement of the linear time-invariant response, non-linear time-invariant response, and 
random and time-variant responses.
Section~\ref{ss:decomposition} describes this decomposition in detail.
In section~\ref{ss:simulation}, we conduct simulations to illustrates the behavior of the proposed FVN-based measurements followed by
the measurement of actual acoustic systems consisting of a microphone and loudspeakers in everyday life situations.
In sectopm~\ref{ss:realtimetool}, we introduce a realtime interactive tool that implements the proposed method.
The tool is accessible as an open-source in our GitHub repository\cite{kawahara2020git}.
Finally, we discuss issues and the prospective applications of the proposed FVN-based methods to other fields.

\section{Background}
The impulse response defines the behavior of a linear time-invariant system.
In the real world, it is impossible to generate the ideal impulse to test the system, and this strategy is not applicable to test real systems.
Various types of TSPs provide a means to circumvent this difficulty.
Swept-Sine and MLS (Maximum Length Sequence) are representative types of TSPs\cite{schroeder1979integrated,aoshima1981jasa,muller2001transfer,guidorzi2015impulse}.
They have temporally spread waveform and have means to compress back to the virtual impulse.
For ideal linear time-invariant systems, using a TSP-based method provides the desired solution for testing target systems.
However, systems in the real world are not linear nor time-invariant.
The background noise and noises in the measuring equipment are unavoidable.
These deviations from the ideal linear time-invariant system introduce measurement errors.
There are several techniques to reduce or separate effects caused by these deviations using post-processing or specialized equipment\cite{dunn1993distortion,farina2000simultaneous,stan2002comparison,burrascano2019swept}.
We introduce a new method to separate the effects of these deviations without introducing specialized equipment or post-processing.
Our method simultaneously derives the linear time-invariant response, non-linear time-invariant response, and 
random and time-variant responses using a set of unit FVNs.
The derivation of the proposed method is lengthy.
However, the resultant procedure is compact, straightforward, and easy to use.

\section{FVN-based response measurements}
This section introduces unit FVN and its use in impulse response measurement first.
Then, we introduce a systematic FVN sequence design method and its application to simultaneous measurement of
the linear time-invariant, the nonlinear time-invariant, and random or time-varying responses.
It is an extension of our previous article\cite{kawahara2019apsipa}.

\subsection{Unit-FVN and response measurement}\label{ss:unitFVN}
A unit-FVN is the impulse response of an all-pass filter.
Only the phase of the transfer function is frequency-dependent.
The following procedure provides the phase\cite{kawahara2018IS,kawahara2019apsipa}.

\subsubsection{Generation of unit FVN}
Similar to the original velvet noise\cite{jarvelainen2007reverberation,valimaki2013ieetr}, the procedure uses two random number sequences $(r_1[n], r_2[n], n \in \mathbb{Z})$ sampled from a uniform distribution in $(0, 1)$.
First, define the center locations ($f_c[n] \in \mathbb{R}$) on the frequency axis of the discrete-time signal using $r_1[n]$.
\vspace{-1mm}
\begin{align}
f_c[n] &= (n - 1 + r_1[n] ) F_d ,
\end{align}%
\vspace{-1mm}%
where $F_d$ represents the average frequency separation of the center locations.

Second, define the coefficient $c_{\varphi}[n]$ of each phase manipulation function using $r_2[n]$.
\vspace{-1mm}
\begin{align}
c_{\varphi}[n] & = (2 \|r_2[n] \|  - 1) \varphi_{\mathrm{max}},
\end{align}
where $\varphi_{\mathrm{max}}$ represents the maximum magnitude of the phase manipulation,
and $\| a \|$ represents the nearest integer of $a$.

The following equation defines the phase $\varphi(f)$ using $f_c[n]$ and $c_{\varphi}[n]$.
\vspace{-1mm}
\begin{align}
\varphi(f) & = \varphi_{+}(f) - \varphi_{-}(f)
, \label{eq:fvnFqd} \\
   \varphi_{+}(f) & = \sum_{n=1}^N c_{\varphi}[n] w_p(f - f_c[n]) ,\\
{}  \varphi_{-}(f)&  = \sum_{n=1}^N c_{\varphi}[n] w_p(f + f_c[n]) 
 = \varphi_{+}(-f) ,
\end{align}
where $N$ represents the number of the allocated phase manipulation function $w_p(f)$.
The function $w_p(f)$ has even symmetry around $f=0$ and has the following form and value.
\vspace{-1mm}
\begin{align}\label{eq:sixTerm}
w_p(f) & = \sum_{m=0}^{5} a_m \cos\left(\frac{m\pi f }{3 c_\mathrm{mag} F_d} \right) ,
\end{align}
where $c_\mathrm{mag}$ represents the time stretching coefficient.
We set $c_\mathrm{mag} = 1$ for FVN generation here.
The coefficients $\{a_m\}_{m=0}^{5}$ given by the following equation provides the optimum attenuation of the maximum sidelobe level\cite{kawahara2017interspeechGS,nattall1981ieee}. 
\vspace{-1mm}
{\small
\begin{align}\label{eq:sixTermCf}
\!\!\! \{a_m\}_{m=0}^{5} & = \left\{
0.2624710164, 0.4265335164, 0.2250165621, \right. \nonumber \\
 & \!\!\!\!\!\!\left. 0.0726831633, 0.0125124215, 0.0007833203 \right\} ,
\end{align}}
A set of numerical optimization defined $\varphi_{\mathrm{max}} = \pi/4$ and made $F_d$ the single design parameter of the unit-FVN.

Inverse discrete Fourier transform of $\exp(j\varphi(f))$ provides the discrete-time signal $h_\mathrm{fvn}[n]$.
It is unit-FVN.
The envelope of a unit-FVN approximates Gaussian distribution and localized in the time domain.
The desired duration, $\sigma_T$ (s), determines the design parameter $F_d = 1 /5\sigma_T $ (Hz).

\subsubsection{Impulse response measurement using a unit FVN}
\begin{figure}[tbp]
\begin{center}
\hfill\includegraphics[width=0.88\hsize]{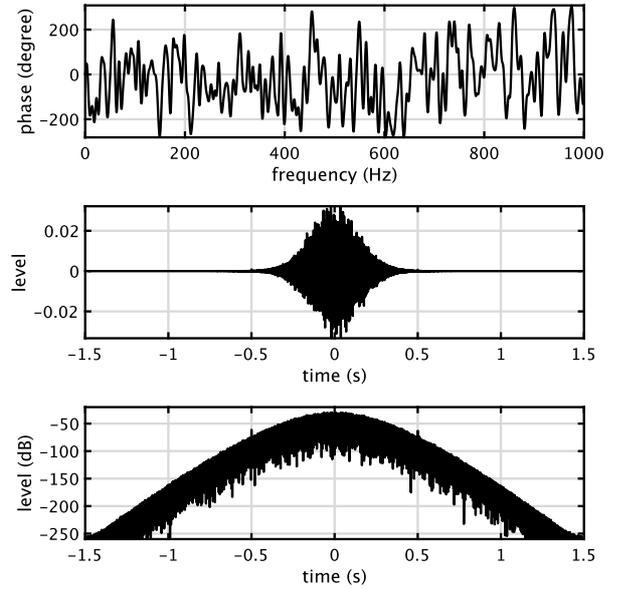}\\
\vspace{3mm}
\caption{Example of a unit FVN. Sampling frequency: 44100~Hz, and the duration $\sigma_\mathrm{T}=0.1$ (s).
The top plot shows the phase $\varphi(f)$ in 0 to 1000~Hz region. The middle plot shows the waveform of a unit FVN $h_\mathrm{fvn}[n]$.
The bottom plot shows the absolute value of the waveform using dB.}
\label{fig:sampleUnitFvn}
\end{center}
\vspace{-4mm}
\end{figure}
Figure~\ref{fig:sampleUnitFvn} shows an example of a unit FVN with the desired duration $\sigma_\mathrm{T}=0.1$ (s) for 44100~Hz sampling frequency.
The bottom plot shows a highly localized amplitude distribution of the unit FVN.
The absolute value of the frequency transfer function of the unit FVN is one because only its phase is frequency-dependent.
Therefore, the convolution of $h_\mathrm{fvn}[n]$ and $h_\mathrm{fvn}[-n]$ yields a unit impulse.
This pulse recovery is the behavior that makes TSP signals useful for impulse response measurement\cite{schroeder1979integrated,aoshima1981jasa,muller2001transfer,guidorzi2015impulse}.

The unique and useful feature of the unit FVN as a TSP is its complexity.
The average frequency distance $F_d$ of the example in Fig.~\ref{fig:sampleUnitFvn} is 2~Hz.
It divides the positive part of the frequency axis into 22050 segments and
allocates each phase manipulation function $w_p(f)$ using two random numbers.
This complexity makes different unit FVNs almost independent (maximum absolute value of cross-correlation is very small.
Refer to Fig.A.2 of~\cite{kawahara2019apsipa}.).
This small correlation provides the key for generating orthogonal sequences described in the following sections.

\subsection{Two FVN sequences and orthogonalization}\label{ss:twoFVNseq}
Generation of two orthogonal sequences from two unit FVNs illustrates the essential idea.
Generate the first sequence by allocating the first unit FVN $h_\mathrm{fvn}^{(1)}[n]$ on a time axis periodically using a constant interval $n_\mathrm{o}$ samples.
Then, generate the second sequence allocate the second unit FVN $h_\mathrm{fvn}^{(2)}[n]$ on a time axis periodically using the same interval $n_\mathrm{o}$ by
changing the polarity every time.
The following equation shows this procedure.
\begin{align}
s_{2}^{(m)}[n] & = \sum_{k=1}^{K} (-1)^{k(m-1)}h_\mathrm{fvn}^{(m)}[n - k n_\mathrm{o}] ,
\end{align}
where $s_{2}^{(m)}[n]$ represents the $m$-th sequence of the generated 2 sequences.
The constant $K$ represents the repetition (length) of the sequence.

\subsubsection{Pulse recovery}
Make a test signal $x_\mathrm{test}[n]$ by adding these two sequences.
Convolution of the time-reversed version of each unit FVN yields recovered signal $q^{(m)}[n]$ consisting of pulses and noise-like leakage caused by cross-correlation.
\begin{align}
q^{(m)}[n] & = h_\mathrm{fvn}^{(m)}[-n] \ast x_\mathrm{test}[n] ,
\end{align}
where $\ast$ represents convolution.

\begin{figure}[tbp]
\begin{center}
\hfill\includegraphics[width=0.9\hsize]{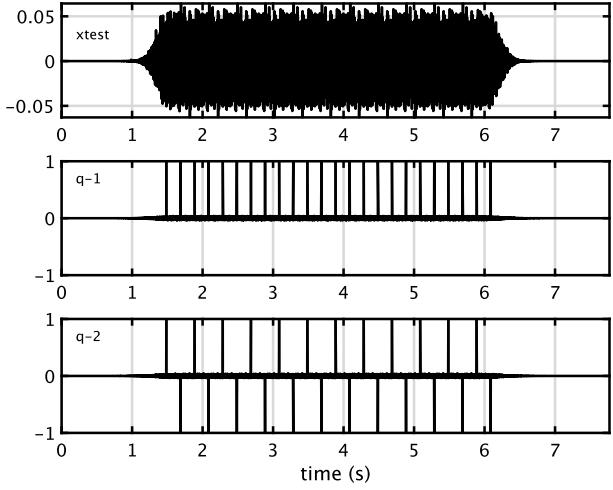}\\
\vspace{-2mm}
\caption{Test signal example and pulse recovered sequences. Top plot shows the test signal $x_\mathrm{test}[n]$.
The middle and bottom plots show the pulse recovered sequences $q^{(m)}[n]$. Note the noise-like signals between recovered pulses.}
\label{fig:recdTwoFVNsequences}
\end{center}
\vspace{-4mm}
\end{figure}
Figure~\ref{fig:recdTwoFVNsequences} shows the test signal $x_\mathrm{test}[n]$ generated using $K=24$ and $n_\mathrm{o}=8820$ samples (0.2~s) and
pulse recovered signals $q^{(m)}[n]$.
Note that there remains a noise-like component between recovered pulses.
It is the result of the cross-correlation between different unit FVNs.
The following step is to cancel out this leakage.

\subsubsection{Orthogonalization}
Note that the pulses in the second recovered sequence are alternating every time.
The leakage from the second sequence in the first sequence is also alternating in each repetition.
Consequently, by adding the time-shifted ($n_\mathrm{o}$ shift) version of the first sequence and the original first sequence, the leakage cancels out.
Similarly, by adding the time-shifted and polarity altered version of the second sequence and the original second sequence, the leakage cancels out.
The following equations show this procedure.
\begin{align}
r_\mathrm{orth}^{(1)}[n] & = \frac{1}{2} (q^{(1)}[n] + q^{(1)}[n - n_\mathrm{o}]) \\
r_\mathrm{orth}^{(2)}[n] & = \frac{1}{2} (q^{(2)}[n] - q^{(2)}[n - n_\mathrm{o}]) ,
\end{align}
where $r_\mathrm{orth}^{(m)}[n]$ represents the $m$-th orthogonalized sequence.

\begin{figure}[tbp]
\begin{center}
\hfill\includegraphics[width=0.9\hsize]{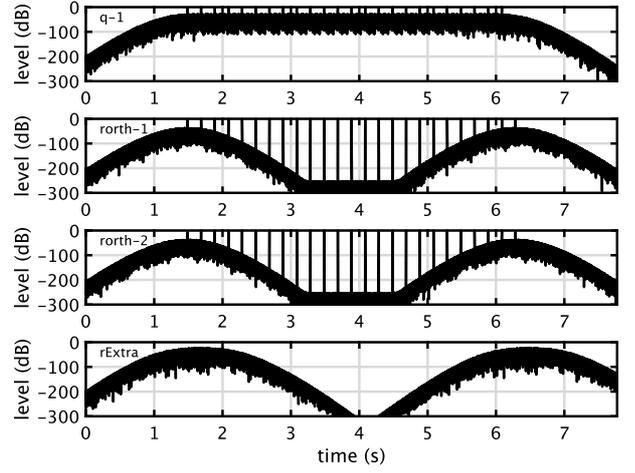}\\
\vspace{-0mm}
\caption{Pulse recovered signal and the orthogonalized signals. The vertical axis represents the level using dB.
The top plot shows the recovered signal $q^{(1)}[n]$. The second and the third panel show the orthogonalized signals $r_\mathrm{orth}^{(m)}[n]$.
The bottom plot shows the extra response $r_\mathrm{extra}^{(X)}[n]$.}
\label{fig:orthTwoFVNsequences}
\end{center}
\vspace{-4mm}
\end{figure}
Figure~\ref{fig:orthTwoFVNsequences} shows the pulse recovered sequence $q^{(1)}[n]$ and the orthogonalized sequences $r_\mathrm{orth}^{(m)}[n]$.
The vertical axis uses dB to illustrate the effect of cancelation clearly.
Note that the canceled leakage level around -260~dB corresponds to $2^{-43}$ and represents numerical operation error.

\subsubsection{Cancelation of the test signal}
Weighted averaging using the weight sequence $\{b_k\}_{k=0}^{3} =\{1, 1, -1, -1\}$, and the time-shifted versions of the convolution of the test signal and other unit FVN removes the effect of the test signal completely.
\begin{align}
r_\mathrm{extra}^{(X)}[n] & = \frac{1}{4}\sum_{k=0}^{3} b_k q^{(X)}[n-k n_\mathrm{o}] ,
\end{align}
where $X$ represents the identifier of the unit FVN used for convolution.
The bottom plot of Fig.~\ref{fig:orthTwoFVNsequences} shows this extra signal.
Note that the extra signal is effectively zero around 4~s.
For practical applications, where lower than -100~dB is negligible, the extra signal is effectively zero from 2.8~s to 5.4~s.
This behavior is useful because suppression of the response due to the linear time-invariant system is possible without extra filter or processing.

\subsection{Four FVN sequences and response measurement}\label{ss:fvnMes4seq}
Similar orthogonalization applies to an arbitrary number of FVN sequences\cite{kawahara2019apsipa}.
For practical applications, using four FVN sequences is especially useful.
This section describes the response measurement procedure using four FVN sequences.

\subsubsection{Generation of four FVN sequences}
Define a weight matrix $\mathrm{B}_4$, which consists of four orthogonal binary rows shown below.
\begin{align}
\mathrm{B}_4 & = \left[ \begin{array}{rrrrrrrr}
1 & 1 & 1 & 1 & 1 & 1 & 1 & 1 \cr
1 & -1 & 1 & -1 & 1 & -1 & 1 & -1 \cr
1 & 1 & -1 & -1 & 1 & 1 & -1 & -1 \cr
1 & 1 & 1 & 1 & -1 & -1 & -1 & -1
\end{array}
\right] \nonumber \\
& = [\mathbf{b}^{(1)}  \mathbf{b}^{(2)} \mathbf{b}^{(3)} \mathbf{b}^{(4)}]^T . \label{eq:orthMat}
\end{align}

The following equation defines the $m$-th FVN sequence $s_4^{(m)}[n], (m = 1, \ldots, 4)$.
\begin{align}
s_4^{(m)}[n] & = \sum_{k=0}^{K -1} h_\mathrm{fvn}^{(m)}[n - kn_\mathrm{o}]b^{(m)}_{\mathrm{mod}(k, 8) + 1} \ ,
\end{align}
where $b_j^{(m)}$ represents the $j$-th element of $\mathbf{b}^{(m)}$, and $K$ represents the number of repetitions.
Note that 8 is the number of columns of $\mathrm{B}_4$.

\subsubsection{Pulse recovery}
Make a test signal $x_\mathrm{test}[n]$ by adding four FVN sequences and
another test signal $x_\mathrm{testR}[n]$ by adding the first three FVN sequences.
\begin{align}
x_\mathrm{test}[n] & = \sum_{m=1}^{4}s_4^{(m)}[n] \\
x_\mathrm{testR}[n] & = \sum_{m=1}^{3}s_4^{(m)}[n] .
\end{align}



Convolution of the test signals and the time-reversed unit-FVNs yields compressed signals $q^{(m)}[n]$ and
$q^{(m)}_\mathrm{R}[n]$.
They consist of recovered pulses corresponding to repetitions.
Since a set of unit-FVNs are not orthogonal, correlations remain as background noise.
\begin{align}\label{eq:compress}
q^{(m)}[n] & = h_\mathrm{fvn}^{(m)}[-n] \ast x_\mathrm{test}[n] \\
q^{(m)}_\mathrm{R}[n] & = h_\mathrm{fvn}^{(m)}[-n] \ast x_\mathrm{testR}[n] ,
\end{align}
where ``$\ast$'' represents the convolution operation.

\begin{figure}[tbp]
\begin{center}
\hfill\includegraphics[width=0.9\hsize]{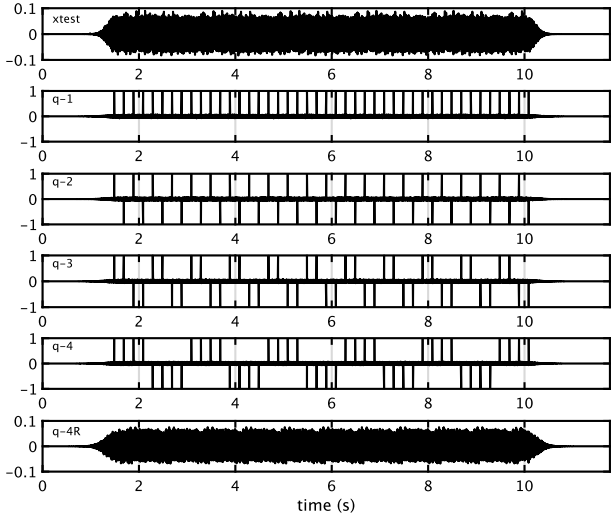}\\
\vspace{2mm}
\caption{Test signal $x_\mathrm{test}[n]$ (top panel), and the pulse compressed signals $q^{(1)}[n], \ldots, q^{(4)}[n], \mbox{and} \ q^{(4)}_\mathrm{R}[n]$ (second to the bottom panels).}
\label{fig:testSigs}
\end{center}
\vspace{-4mm}
\end{figure}
Figure~\ref{fig:testSigs} shows $q^{(1)}[n], \ldots, q^{(4)}[n]$, and $q^{(4)}_\mathrm{R}[n]$ in the second to the bottom panels.
The number of repetitions is $K=44$ and $n_\mathrm{o}=8820$ (0.2~s for 44100~Hz sampling).
The duration of each unit FVN is $\sigma_\mathrm{T}=0.1$~s.

The cross-correlation with the extra components due to the system nonlinearity and the system and measurement noise also remains in this signal because they are not orthogonal to any unit-FVNs.
That is the bottom panel of Fig.~\ref{fig:testSigs} that shows $q^{(4)}_\mathrm{R}[n]$.
Note that convolution using a time-reversed FVN does not modify the extra component's power because it is the impulse response of an all-pass filter.

\subsubsection{Orthogonalization}
The inner product of $\mathbf{b}^{(i)}$ and $\mathbf{b}^{(j)}$ of (\ref{eq:orthMat}) is zero when $i\ne j$.
Periodic time shift and averaging using the binary weights in $\mathbf{b}^{(1)}, \ldots, \mathbf{b}^{(4)}$ remove these cross-correlations.
The following equation yields the orthogonalized signal $r_\mathrm{itr}^{(m)}[n]$.
\begin{align}\label{eq:orthogonal}
r_\mathrm{itr}^{(m)}[n] & = \frac{1}{8}\sum_{k=0}^{8 - 1} q^{(m)}[n - kn_\mathrm{o}]b^{(m)}_{\widetilde{k+1}} \ ,
\end{align}
where $8$ is the length of the weight vectors $\mathbf{b}^{(1)}, \ldots \mathbf{b}^{(4)}$.
The notation $\widetilde{k+1}$ represents cyclic indexing between 1 and 8.

\begin{figure}[tbp]
\begin{center}~\\
~\hfill\includegraphics[width=0.9\hsize]{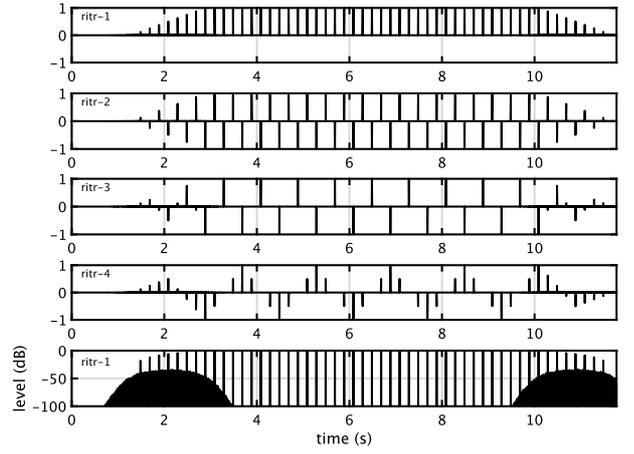}\\
\vspace{1mm}
\caption{Orthogonalized signals $r_\mathrm{itr}^{(1)}[n], \ldots, r_\mathrm{itr}^{(4)}[n]$ (from top to the fourth panel).
The bottom panel shows $\left| r_\mathrm{itr}^{(1)}[n]\right|$ using dB vertical axis.}
\label{fig:orthognalTestSignal}
\end{center}
\vspace{-6mm}
\end{figure}
The top four panels of Fig.~\ref{fig:orthognalTestSignal} shows the orthogonalized signals $r_\mathrm{itr}^{(1)}[n], \ldots, r_\mathrm{itr}^{(4)}[n]$.
The bottom panel of Fig.~\ref{fig:orthognalTestSignal} shows the absolute value of  $r_\mathrm{itr}^{(1)}[n]$ using dB scale.
Noise-like cross-correlation from 3.5~s to 9~s is completely cancelled.

\subsubsection{Synchronous averaging for unit impulse recovery}
Assume that the observed test signal consists of additive random noise.
Then, the orthogonalized signal $r_\mathrm{itr}^{(m)}[n]$ made from the observed test signal also consists of the noise.
Because the orthogonalization process consists of time-shifting eight times and averaging
two segments of length $n_\mathrm{o}$ separated by $8n_\mathrm{o}$ samples are mutually independent.
The following synchronous averaging reduces the effect of random fluctuation.
\begin{align}\label{eq:syncAv}
r^{(m)}[n] & = \frac{1}{\#(\Omega)} \sum_{n_\mathrm{ini} + 8 k n_\mathrm{o} \in \Omega} r_\mathrm{itr}^{(m)}[n + n_\mathrm{ini} + 8 k n_\mathrm{o}] ,
\end{align}
where $0 \le n < n_\mathrm{o}$, and the symbol $\Omega$ represents the region where cross-correlation is effectively vanished.
The function $\#(\Omega)$ yields the number of pulses separated by $8n_\mathrm{o}$ and located inside the region.
Then, further averaging four responses $r^{(m)}[n], (m = 1, \ldots, 4)$ provides the final averaged response $r[n]$.
\begin{align}
r[n] & = \frac{1}{4} \left(r^{(1)}[n] + r^{(2)}[n] + r^{(3)}[n] + r^{(4)}[n] \right) .
\end{align}

The next section introduces a procedure to expand the segment length of the impulse.
It uses the alternate test signal $x_\mathrm{testR}[n]$.
The averaged response for the test signal $r_\mathrm{R}[n]$ is the average of the first three FVN sequences.
\begin{align}\label{eq:oneAvResp}
r_\mathrm{R}[n] & = \frac{1}{3} \left(r^{(1)}_\mathrm{R}[n] + r^{(2)}_\mathrm{R}[n] + r^{(3)}_\mathrm{R}[n] \right) ,
\end{align}
where $r^{(m)}_\mathrm{R}[n]$ is the counter part of $r^{(m)}[n]$ in (\ref{eq:syncAv}) and
uses $r_\mathrm{itrR}^{(m)}[n]$.
The orthogonalized signal $r_\mathrm{itrR}^{(m)}[n]$ uses $q^{(m)}_\mathrm{R}[n]$ instead of $q^{(m)}[n]$ in the counter part of (\ref{eq:orthogonal}).

The whole procedure described applies to the impulse response measurement of the target system.
For impulse response measurement, instead of using the test signal, apply the procedure to the output of the target system to the test signal input. 

\subsubsection{Expansion of the segment length}\label{ss:respExp}
The orthogonalized signals $r^{(3)}[n]$ and $r^{(3)}[n]$ are periodic signal with the period $8n_\mathrm{o}$ as shown in Fig.~\ref{fig:orthognalTestSignal} from 3.5~s to 9~s.
Combining these orthogonalized signals makes it possible to derive a segment with the length $4n_\mathrm{o}$ consisting of one impulse at the
beginning.
This segment expansion is useful for applications where only a short repetition period is acceptable.

Make a $4 \times 8$ matrix $\mathrm{A}$ with element $a_{m, j}$ defined by the following equation where $M=4$.
\begin{align}
a_{m, j} & = \frac{1}{2^{M-1}}\sum_{k=0}^{2^{M-1} - 1} b^{(m)}_{\widetilde{k+1}}b^{(m)}_{\widetilde{j-k+1}} 
\end{align}
\begin{align}
\!\!\!\!\!\mathrm{A} & = 
\!\frac{1}{8}\!\left[
\begin{array}{rrrrrrrr}
 \! \!   8  &\!   8  & \  8  &\!   8  & \  8  &\!   8   &\!  8  &\!   8 \\
 \! \!   8  &\!  -8  & \  8  &\!  -8  & \  8  &\!  -8   &\!  8  &\!  -8 \\
 \!  \!  0  &\!   8  & \  0  &\!  -8  & \  0  &\!   8   &\!  0  &\!  -8 \\
 \!  \! -4  &\!   0  & \  4  &\!   8  & \  4  &\!   0   &\! -4  &\!  -8
\end{array}
\right] .
\end{align}

Solving the following equation provides the coefficients to combine the orthogonalized signals to yield the desired periodic segment of length $4n_\mathrm{o}$.
\begin{align}
\mathbf{v} & = \mathrm{A}^T\mathbf{c} ,
\end{align}
where $\mathbf{c}$ consists of the coefficients for the combination, and $\mathbf{v}$ represents a four-element vector having one non-zero element.
An exhaustive search of four possible cases provides the following two solutions.
\begin{align}
\mathbf{c}_1 & = \frac{1}{4}\left[
\begin{array}{r}
 1 \\
 -1 \\
 2 \\
 0   
\end{array}
\right], 
\ & \mathbf{c}_2 = \frac{1}{4}\left[
\begin{array}{r}
 1 \\
 -1 \\
 -2 \\
 0   
\end{array}
\right] ,
\end{align}
where both solutions set zero weight for the fourth orthogonalized signal.
Then, the following equation provides the signal $r_\mathrm{Xtmp}[n]$ with the expanded periodic segment length $4n_\mathrm{o}$.
\begin{align}\label{eq:orthSeq}
r_\mathrm{Xtmp}[n] & = \frac{1}{4}\left(r_\mathrm{itrR}^{(1)}[n] - r_\mathrm{itrR}^{(2)}[n] + 2 r_\mathrm{itrR}^{(3)}[n]\right) ,
\end{align}
where we use $r_\mathrm{itrR}^{(m)}[n]$ instead of using $r_\mathrm{itr}^{(m)}[n]$ because there is no need to include the fourth FVN sequence in the test signal.
Afterwards, we use $x_\mathrm{testR}[n]$ instead of $x_\mathrm{test}[n]$.

\begin{figure}[tbp]
\begin{center}
\includegraphics[width=0.92\hsize]{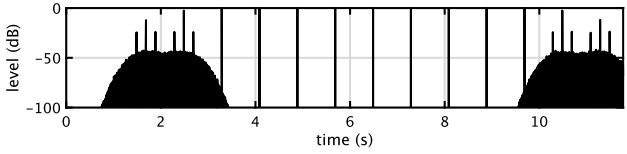}\\
\vspace{-2mm}
\caption{The signal $r_\mathrm{Xtmp}[n]$ with the expanded periodic segment. The vertical axis uses dB.}
\label{fig:orthogSigs}
\end{center}
\vspace{-5mm}
\end{figure}
Figure~\ref{fig:orthogSigs} shows this signal $r_\mathrm{Xtmp}[n]$.
The periodic segments consisting of pulses from 4~s to 8.5~s are free of cross-correlation.
Finally, the synchronous averaging provides the expanded unit signal $r_\mathrm{XPD}[n]$ with length $4n_\mathrm{o}$.
\begin{align}\label{eq:respXPDAv}
r_\mathrm{XPD}[n] & = \frac{1}{\#(\Omega_\mathrm{XPD})}\sum_{n_k \in \Omega_\mathrm{XPD}} r_\mathrm{Xtmp}[n + n_k] ,
\end{align}
where $\Omega_\mathrm{XPD}$ represents the set of recovered pulse locations inside the free-of-cross-correlation region (for example, in Fig.~\ref{fig:orthogSigs}, from 4~s to 8.5~s.).
The function $\#(\Omega_\mathrm{XPD})$ provides the number of pulses in the set $\Omega_\mathrm{XPD}$.

\begin{figure}[tbp]
\begin{center}
\includegraphics[width=0.92\hsize]{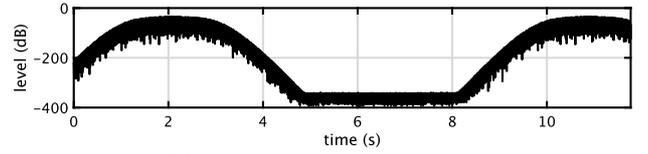}\\
\vspace{-2mm}
\caption{The signal $r_\mathrm{itrR}^{(4)}[n]$ for the test signal $x_\mathrm{testR}[n]$. The vertical axis uses dB.}
\label{fig:theFourthChannel}
\end{center}
\vspace{-5mm}
\end{figure}
Figure~\ref{fig:theFourthChannel} shows the orthogonalized signal using the fourth unit FVN using the test signal $x_\mathrm{testR}[n]$, which does not consist of the fourth unit FVN.
Note that no component due to $x_\mathrm{testR}[n]$ appears from 4.6~s to 8~s.
This behavior is useful for measuring observation noise while measurement described later.

The next possible segment length expansion is using eight FVN sequences.
We do not discuss it in detail here because the number of repetitions required for the expansion is $2^7=128$ times, and we cannot find useful applications.

\subsubsection{Summary for liner time-invariant systems}
For linear time-invariant systems, applying the test signal $x_\mathrm{testR}[n]$ and processing the output of the system using the procedure mentioned above provides two impulse responses; $r_\mathrm{R}[n]$ and $r_\mathrm{XPD}[n]$.
The former has higher SNR for random observation noise.
The selection of these alternatives is a trade-off between SNR and response length and depends on the application.
The orthogonalization procedure using the fourth unit-FVN removes all cross-correlation effects and leaves components due to observation noise while applying the test signal $x_\mathrm{testR}[n]$.
The integration of all these procedures enables the simultaneous measurement of the impulse response and SNR.
It also provides simultaneous measurement of the nonlinear time-invariant component.
These descriptions summarize the proposed method for measuring linear time-invariant systems.
The next section generalizes this method for measuring nonlinear systems.

\subsection{Measuring deviations from linear time-invariant systems}\label{ss:decomposition}
We consider three sources of deviations from linear time-invariant systems encountered in measuring acoustic systems.
First is observation noise such as background noise, microphone and amplifier noise, noise from the power line, and other noise sources.
The second is nonlinearity, such as loudspeaker nonlinearity. 
The third is temporal variation, such as sound speed variation due to airflow and temporal change, and clock difference of the AD converter and the DA converter.
When the acoustic systems are consisting of digital transmission, compression algorithms in the system introduce deviations that mixing all three deviations.

\subsubsection{Measuring random (and time varying) component}
Figure~\ref{fig:theFourthChannel} shows that the orthogonalized signal using the fourth unit FVN $r_\mathrm{itrR}^{(m)}[n]$ does not consist of the test signal $x_\mathrm{testR}[n]$ related component.
Then, applying the synchronous averaging process to $x_\mathrm{testR}[n]$ yields the signal $r_\mathrm{RV}[n]$ which consisting of the averaged extra components.
\begin{align}\label{eq:rrandNoise}
r_\mathrm{RV}[n] & = \frac{1}{L + 1}\sum_{k=0}^{L}r_\mathrm{itrR}^{(4)}[n + 8kn_\mathrm{o} + n_0] ,
\end{align}
where $L$ is determined to make $n_0$ and $n_0+(8L+1)n_\mathrm{o}$ are inside the region of no cross correlation.
For example, using Fig.~\ref{fig:theFourthChannel} as an example, 4.6~s to 8~s is ideal, and 3.5~s to 9~s is practically acceptable.
Note that all possible combinations of the first three FVN sequences repeat in $4n_\mathrm{o}$, and the orthogonalization process averages two $4n_\mathrm{o}$ periods using the opposite sign for calculating $r_\mathrm{itrR}^{(4)}[n]$.
It cancels the effects of time-invariant nonlinearity.
In other words, $r_\mathrm{RV}[n]$ consists of component due to random (and/or time-varying) component.

When the measurement only consists of a stable random component having variance $\sigma_\mathrm{Ro}^2$, the following equation provides the relation between the variance and the sample variance $\sigma_\mathrm{RV}^2$ of the fourth signal $r_\mathrm{RV}[n]$.
\begin{align}\label{eq:NVrec}
\!\!8(L+1)\sigma_\mathrm{Ro}^2 & = \sigma_\mathrm{RV}^2 
= \!\frac{1}{8n_\mathrm{o} - 1}\!\!\sum_{n=0}^{8n_\mathrm{o} - 1}\! \!\!\left|r_\mathrm{RV}[n] - \overline{r_\mathrm{RV}} \right|^2\! ,
\end{align}
where $\overline{r_\mathrm{RV}}$ represents the average of $r_\mathrm{RV}[n]$ in terms of $n$.

\subsubsection{Measuring nonlinear time-invariant component}
We assume the following conditions because each unit FVN has a high degree of freedom and mutually independent.
\begin{enumerate}
\item The component due to nonlinearity does not correlate with each unit FVN consisting of the test signal.
\item The component yielded by the combination of FVN sequences are time-invariant. The same combination yields the same component.
\end{enumerate}

The orthogonalization procedure of $r_\mathrm{itrR}^{(m)}[n], (m \ne 4)$ is identical for the two $4n_\mathrm{o}$-length segments.
Therefore the sample squared deviation is $1/2^2$ times of the actual deviation.
Consider the first three responses $r_\mathrm{R}^{(1)}[n], r_\mathrm{R}^{(2)}[n], \mbox{and} \ r_\mathrm{R}^{(3)}[n], $ to $x_\mathrm{testR}[n]$.
\begin{align}
r_\mathrm{R}^{(m)}[n] & = \frac{1}{8(L+1)} \sum_{k=0}^L r_\mathrm{itrR}^{(m)}[n + 8kn_\mathrm{o} + n_0] .
\end{align}

This averaging process does not change nonlinearity contribution because each repetition of $L+1$ has an equal contribution.
The following equation defines the sample variance ($\sigma_\mathrm{N}^2$) in terms of FVN sequences.
\begin{align}
\sigma_\mathrm{N}^2 & = \frac{9}{\#(\Omega_\mathrm{seg})} \sum_{m=1}^{3}\sum_{\ n \in \Omega_\mathrm{seg}} \left|d^{(m)}_\mathrm{R}[n] \right|^2 \\
d^{(m)}_\mathrm{R}[n] & = r^{(m)}_\mathrm{R}[n] - \overline{r_\mathrm{R}[n]} \label{eq:devFromLTI} \\
\overline{r_\mathrm{R}[n]} & = \frac{1}{3} \sum_{m=1}^{3} r_\mathrm{R}^{(m)}[n] ,
\end{align}
where the sample variance of $d_\mathrm{R}^{m}[n]$ uses $1/(3-1)$ for normalization
and consequently $4\cdot 3/ 2 = 6$ times yields the proper value.
The average of each power spectrum of $d^{(m)}_\mathrm{R}[n]$ provides the spectral contribution of the nonlinear component.
Also, the power spectrum of $r_\mathrm{RV}[n]$ provides the spectral contribution of the random (and time-varying) component.

\subsubsection{Summary of measurement with nonlinearity and extra component}
Using test signal $x_\mathrm{testR}[n]$ to the target system and get output of the system $y_\mathrm{testR}[n]$.
Applying procedure described above and get two responses $r_\mathrm{R}[n]$ using (\ref{eq:oneAvResp}) for higher-SNR and
$r_\mathrm{XPD}[n]$ using (\ref{eq:respXPDAv}) for expanded response length.
For observation noise and extra contribution (such as time varying response), calculate $r_\mathrm{RV}[n]$ using (\ref{eq:rrandNoise}) and calibrate it using (\ref{eq:NVrec}).
For nonlinear time-independent contribution, calculate each deviation $d_\mathrm{R}^{(m)}[n]$ using (\ref{eq:devFromLTI}) and calculate RMS (root mean squared) average of each power spectra.
In addition to these, the initial or final low-amplitude region of the output signal $y_\mathrm{testR}$ of the system,
directly provides the background noise level when the background noise is stationary.

\section{Neumerical examples}\label{ss:simulation}
First, this section introduces the behavior of the proposed procedure by simulations using know mode.
Then, several measurements using real acoustic systems provide examples in actual applications.

\subsection{Simulations}
We use a simulated loudspeaker consisting of the linear time-invariant response and instantaneous nonlinearity to input.
\begin{figure}[tbp]
\begin{center}
\hfill\includegraphics[width=0.92\hsize]{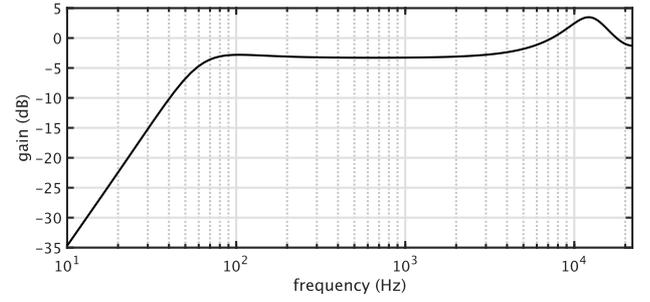}\\
\vspace{-2mm}
\caption{Simulated frequency response.}
\label{fig:simresp}
\end{center}
\vspace{-4mm}
\end{figure}
Figure~\ref{fig:simresp} shows the frequency amplitude response of the simulator.
It has complex conjugate poles at 70~Hz and 8000~Hz with bandwidths 50~Hz and 1200~Hz, respectively.
They model low-frequency cut-offs and small resonance in the higher frequency region.
In addition to these poles, additional second-order differentiation makes the model simulate loudspeakers.

We used the following nonlinearity consisting of asymmetry and saturation behavior.
\begin{align}\label{eq:nlFunction}
f(x) & = \frac{2}{1+ \exp(-2 (x + \alpha \exp(x)))} - 1 - \alpha ,
\end{align}
where $\alpha$ represents the parameter determining asymmetry.
We use 0.3 for the following simulations.
Note that this nonlinearity consists of higher-order components.
\begin{figure}[tbp]
\begin{center}
\hfill\includegraphics[width=0.92\hsize]{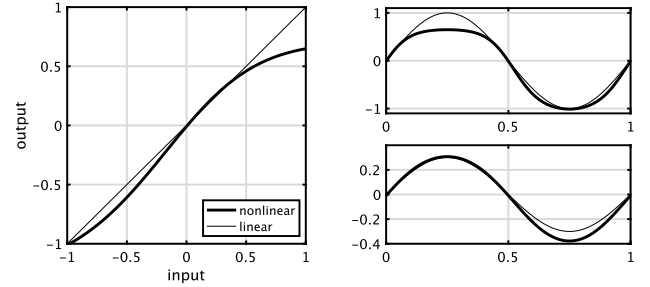}\\
\vspace{-2mm}
\caption{Simulated nonlinearity. The left panel shows the input-to-output relation.
Right two panels show waveform distortion examples for full range input and small ($\pm 0.3$, peak-to-peak) input.}
\label{fig:nodelNl}
\end{center}
\vspace{-4mm}
\end{figure}
Figure~\ref{fig:nodelNl} shows this nonlinearity and examples of waveform distortion.

\subsubsection{Power separation test}
We tested signal power separation into the linear, nonlinear, and random components.
First, we normalized $x_\mathrm{testR}[n]$ by its sample standard deviation and defined its level $0$~dB.
Then, we fed attenuated ($0, -10, \ldots, -50$~dB) versions to the nonlinearity defined by (\ref{eq:nlFunction}) and
calculated power of the linear, the nonlinear, and the random (and extra) component using the procedure mentioned in the previous section.

\begin{figure}[tbp]
\begin{center}
\hfill\includegraphics[width=0.92\hsize]{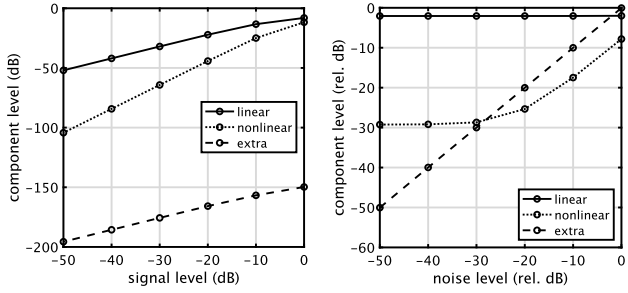}\\
\vspace{-1mm}
\caption{Component power separation test.
The left panel shows the effect of input test signal level on the nonlinear conversion (\ref{eq:nlFunction}).
The right panel shows the effect of additive Gaussian noise level on the output of the nonlinear conversion.
The noise level is relative to the converted output level.}
\label{fig:simOrth}
\end{center}
\end{figure}
The left panel of Fig.~\ref{fig:simOrth} shows the results.
The horizontal axis represents the signal level, and the vertical axis represents the level of the separated component.
In this simulation, we added random Gaussian noise, which is 150~dB lower than the attenuated test signal.
The plot shows that the separation is successful, and the nonlinear component shows a steeper increment slope than the level increment.

Then, we used -25~dB attenuation to set the input signal to the nonlinearity and
added Gaussian noise.
The Gaussian noise level was ($0, -10, \ldots, -50$~dB) to the nonlinear converted output.
The right panel of Fig.~\ref{fig:simOrth} shows the results.
The horizontal axis shows the noise level relative to the nonlinear converted signal level.
The vertical axis shows the separated component level normalized by the nonlinear converted signal level.
Note when the random component has a higher intensity than the nonlinear component, it smears the nonlinear component.

Note that separated levels represent the status at the observation point.
The linear component's final estimation has smaller effects from these nonlinear and extra components because of the synchronous averaging.

\subsubsection{Spectrum separation test}
Then, we conducted acoustic system measurement simulations.
First, we used the nonlinear converted signal and inputted it to the loudspeaker simulator shown in Figure~\ref{fig:simresp}.
We separated into the linear time-invariant response, the nonlinear time-invariant response, and extra component as the background.
We also calculated (raw) background level using the preceding low-amplitude part of the tested signal.
We calibrated using the test signal with -25~dB attenuation as the reference and assumed it to have the sound pressure level 70~dB using A-weighting.
Then, we also used the test signal, 10~dB stranger.

\begin{figure}[tbp]
\begin{center}
\hfill\includegraphics[width=0.9\hsize]{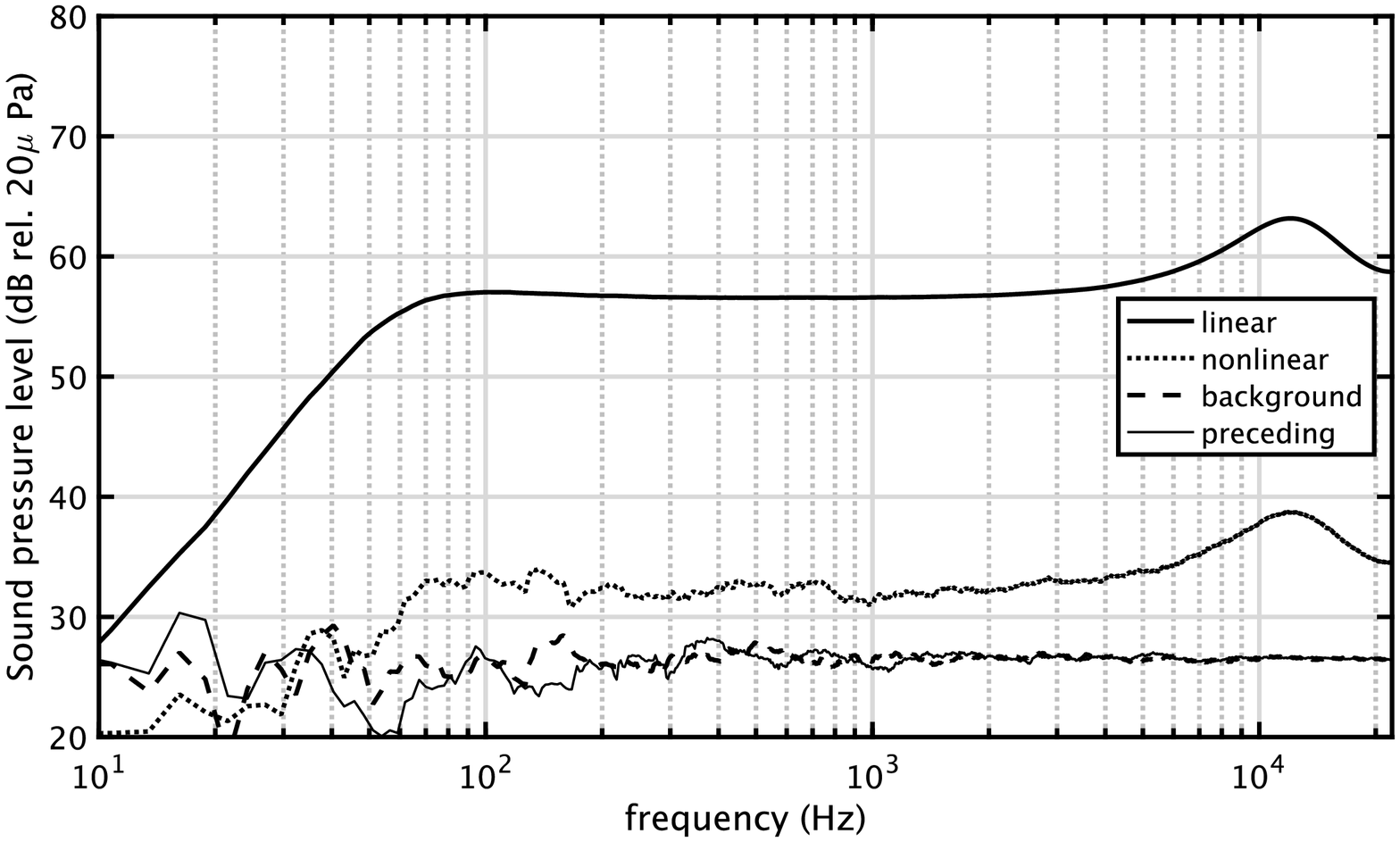}\\
\hfill\includegraphics[width=0.9\hsize]{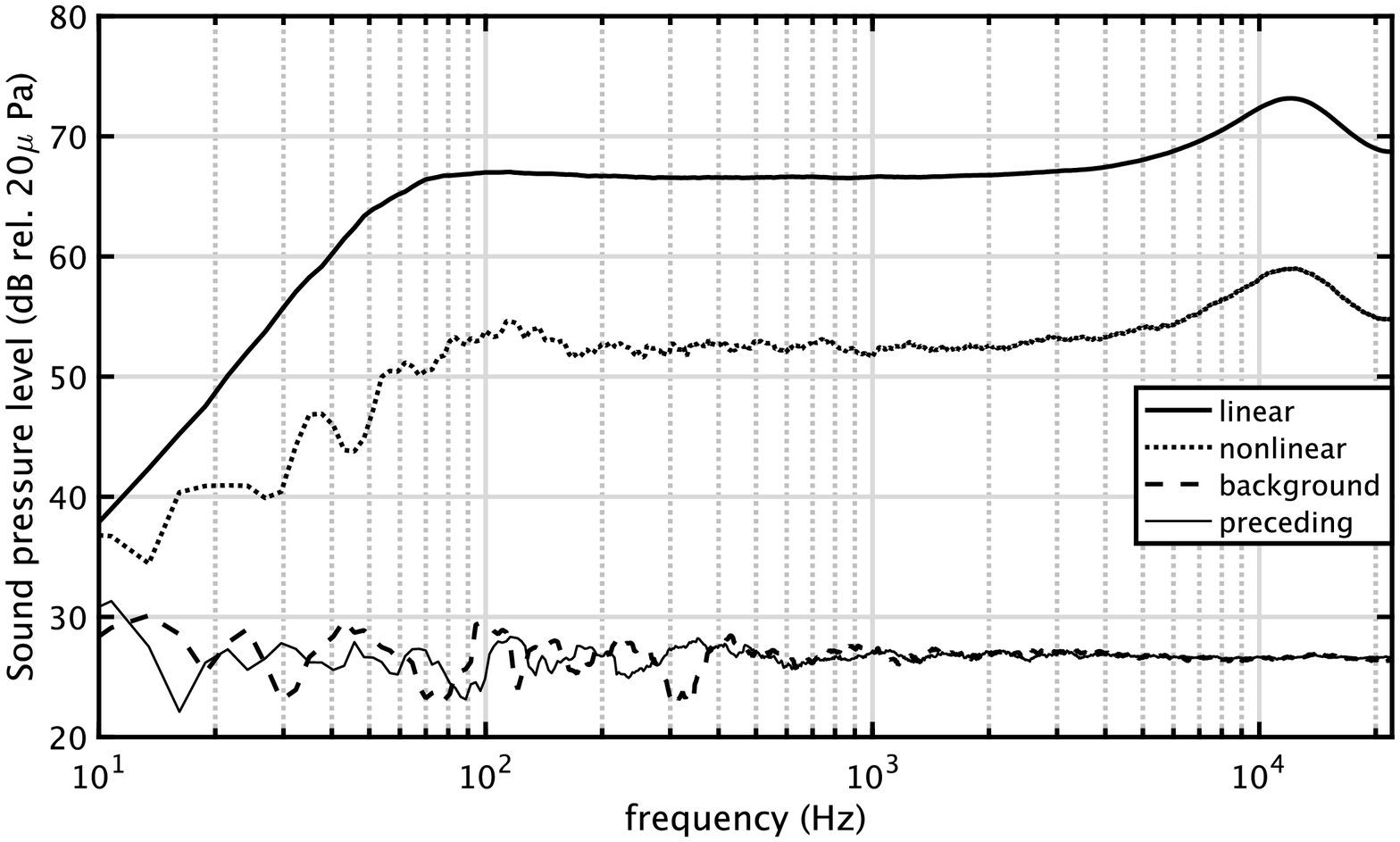}\\
\vspace{-2mm}
\caption{Power spectra of separated components. The vertical axis represents the simulated sound pressure level in a 1/3 octave width.
The level using A-weight is 70~dB for the top panel and 80~dB for the bottom panel.}
\label{fig:simFreqRespWO}
\end{center}
\end{figure}
Figure~\ref{fig:simFreqRespWO} shows the power spectra of the responses.
``linear'' in the legend represents the power spectrum of $r_\mathrm{R}[n]$.
``nonlinear'' represents the RMS value of the power spectra of $d_\mathrm{R}^{(m)}[n]$ with calibration.
``background'' represents the power spectrum of $r_\mathrm{RV}[n]$.
``preceding'' represents the background level using the preceding low-amplitude part of the tested signal.
Figure~\ref{fig:simFreqRespWO} illustrates that decomposition of the deviation from linear time-invariant systems in the frequency domain works.
Note that the nonlinear response increases 20~dB when the input level increased 10~dB.

We used smoothed power spectrum $p_\mathrm{S}(f)$ of the original power spectrum $p(f)$ using 1/3 octave rectangular smoother defined below.
\begin{align}
p_\mathrm{S}(f) = \frac{1}{(2^{\frac{1}{6}} - 2^{-\frac{1}{6}}) f}\int_{f 2^{-\frac{1}{6}}}^{f 2^{\frac{1}{6}}} p(\nu) d\nu .
\end{align}
The legend has an item ``preceding'' representing the (smoothed) power spectrum calculated using the initial low-amplitude region of the response $y[n]$ with added noise.
Note that ``background'' and ``preceding'' are effectively overlaid, suggesting the noise level calibration using (\ref{eq:NVrec}) is valid.

\subsubsection{Response expansion}
The final simulation is the response expansion procedure described in~\ref{ss:respExp}.
We used a simulated impulse response of a large room using Audacity\cite{audacitySoft} and calculated convolution with the simulator output.
\begin{figure}[tbp]
\begin{center}
\hfill\includegraphics[width=0.9\hsize]{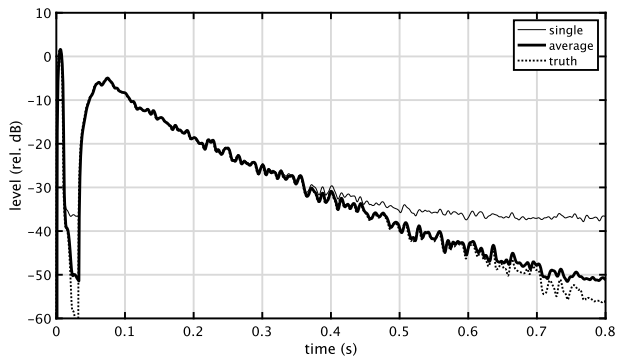}\\
\vspace{-2mm}
\caption{Smoothed power response with simulated reverberation using expanded response $r_\mathrm{XPD}[n]$. The simulated loudspeaker output with 80~dB sound pressure level was the input to the reverberation simulator. Note that the allocation interval of unit FVNs is 0.2~s. The legend ``average'' shows an average of 40 iterations.}
\label{fig:simReverb}
\end{center}
\vspace{-6mm}
\end{figure}
Figure~\ref{fig:simReverb} shows the smoothed (using 10~ms Hanning window) power responses.
The test condition is the same as the bottom plot of Figure~\ref{fig:simFreqRespWO} where the primary deviation from linear time-invariant systems is due to nonlinearity.
We generated a set of unit FVNs 40 times using different random seeds.
The allocation interval of unit FVNs was 0.2~s.

The thick line with legend ``linear'' shows the expanded response $r_\mathrm{XPD}[n]$ defined by (\ref{eq:respXPDAv}).
The thin dotted line with legend ``truth''  shows the ground truth with no observation noise.
Figure~\ref{fig:simReverb} shows that the averaged response provides reliable results up to 6~s, while the result using a single measurement assures up to 3.5~s.
It indicates that the extension works.
Averaging different measurements using different FVN set reduces the effects of nonlinearity inversely proportional to the square root of the number of repetitions.

\subsection{Acoustic system measurement examples}
\begin{figure}[tbp]
\begin{center}
\hfill\includegraphics[width=0.99\hsize]{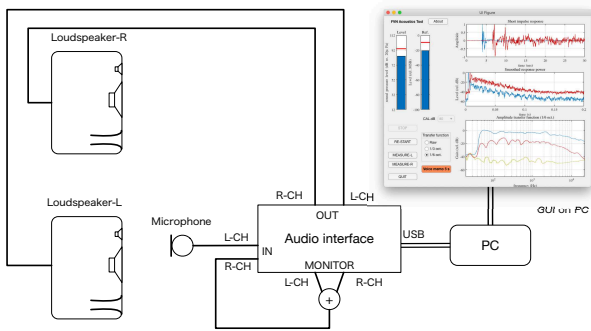}\\
\vspace{-8mm}
\caption{Acoustic measurement setting.}
\label{fig:setting}
\end{center}
\end{figure}
Figure~\ref{fig:setting} shows the acoustic measurement setting of a set of loudspeakers in an ordinary living room of residential area.%
\footnote{Microphone: DPA 4066 Omnidirectional miniature condenser microphone, Audio Interface: Presonus STUDIO 2|6, Loudspeaker: IKmultimedia iLoud Micro Monitor, PC: MacBook Pro 13", Software: MATLAB 2020a Update-4, Sampling frequency: 44100~Hz, Bit per sample: 24.}
All measurements used an interactive and realtime tool for acoustic measurement described in the next section.
The left channel input is the microphone output, and the right channel input is the mixed signal of the stereo phone output.\footnote{%
This right channel input is optional. It is for the response delay measurement.
A standard laptop with a built-in microphone and a stereo sound output can execute this procedure without adding an audio interface.}
The left channel of the loudspeaker to the microphone is the target acoustic system in the following examples.
The distance between the loudspeaker and the microphone is 20~cm.

\begin{figure}[tbp]
\begin{center}
\hfill\includegraphics[width=0.99\hsize]{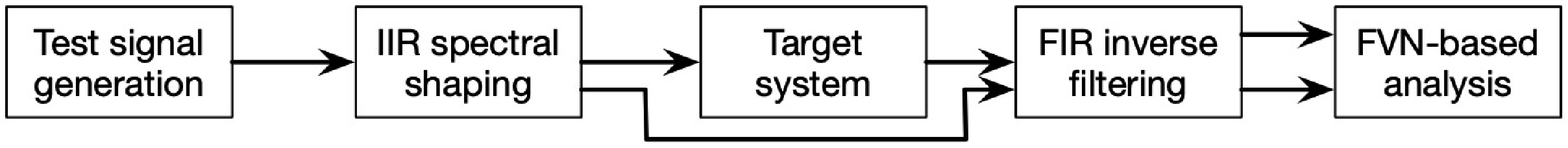}\\
\vspace{-3mm}
\caption{Schematic diagram of the test procedure.}
\label{fig:testProcedure}
\end{center}
\end{figure}
Figure~\ref{fig:testProcedure} shows the test procedure.
In this test, we introduced an IIR spectral shaper and an FIR inverse filter.
The spectral shaping using a 44~tap IIR (Infinite Impulse Response) filter made the test signal has the pink noise spectrum.
The LPC (Linear Predictive Coding\cite{itakura1970ieice,atal1971jasa}) procedure designed the 44-tap IIR filter for pink-noise spectrum shaping.
This LPC-based design procedure of the IIR spectral shaper also applies to design the shaper for optimal SNR measurements\cite{ochiai2013a}.

\begin{figure}[tbp]
\begin{center}
\hfill\includegraphics[width=0.9\hsize]{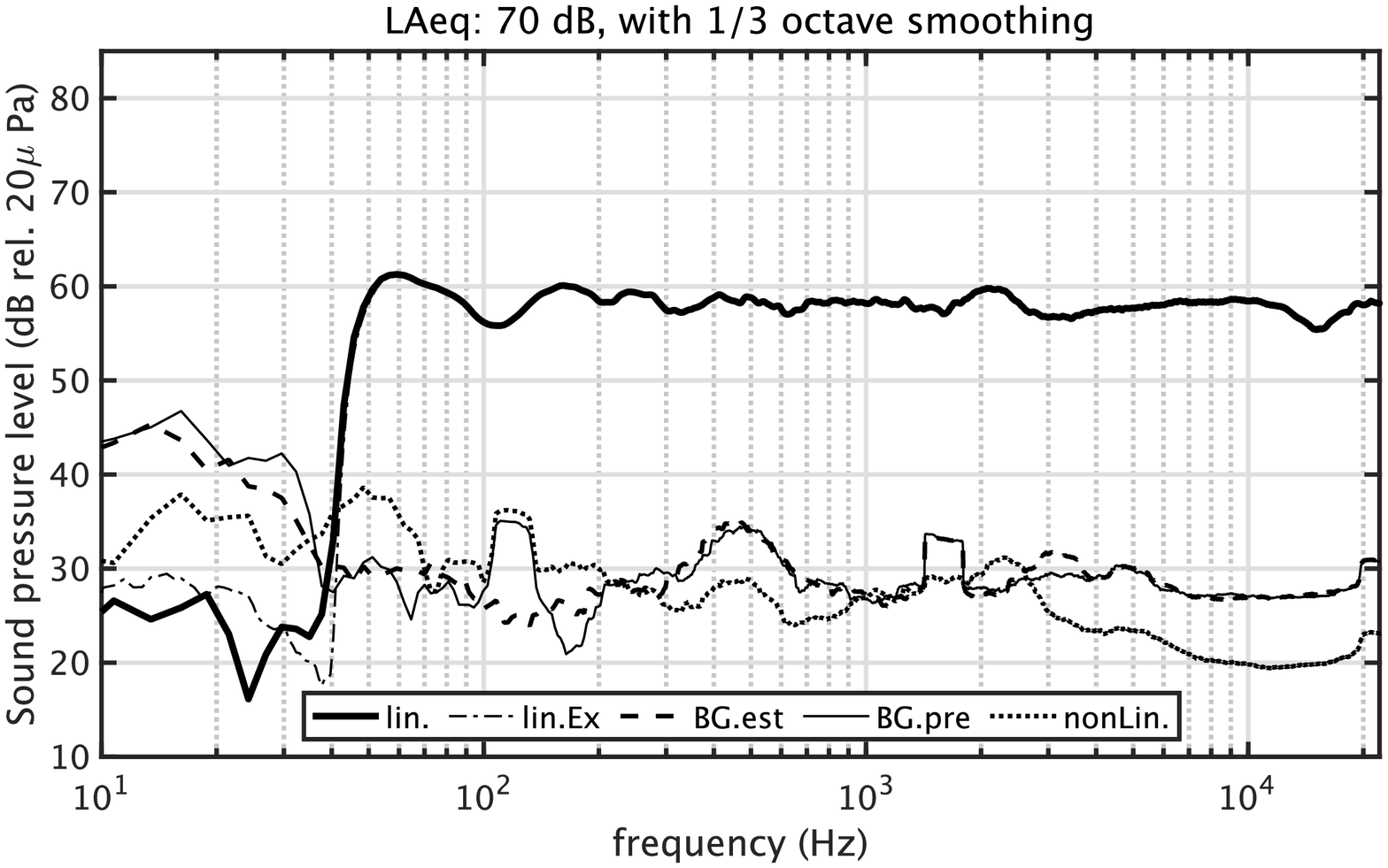}\\
\vspace{2mm}
\hfill\includegraphics[width=0.9\hsize]{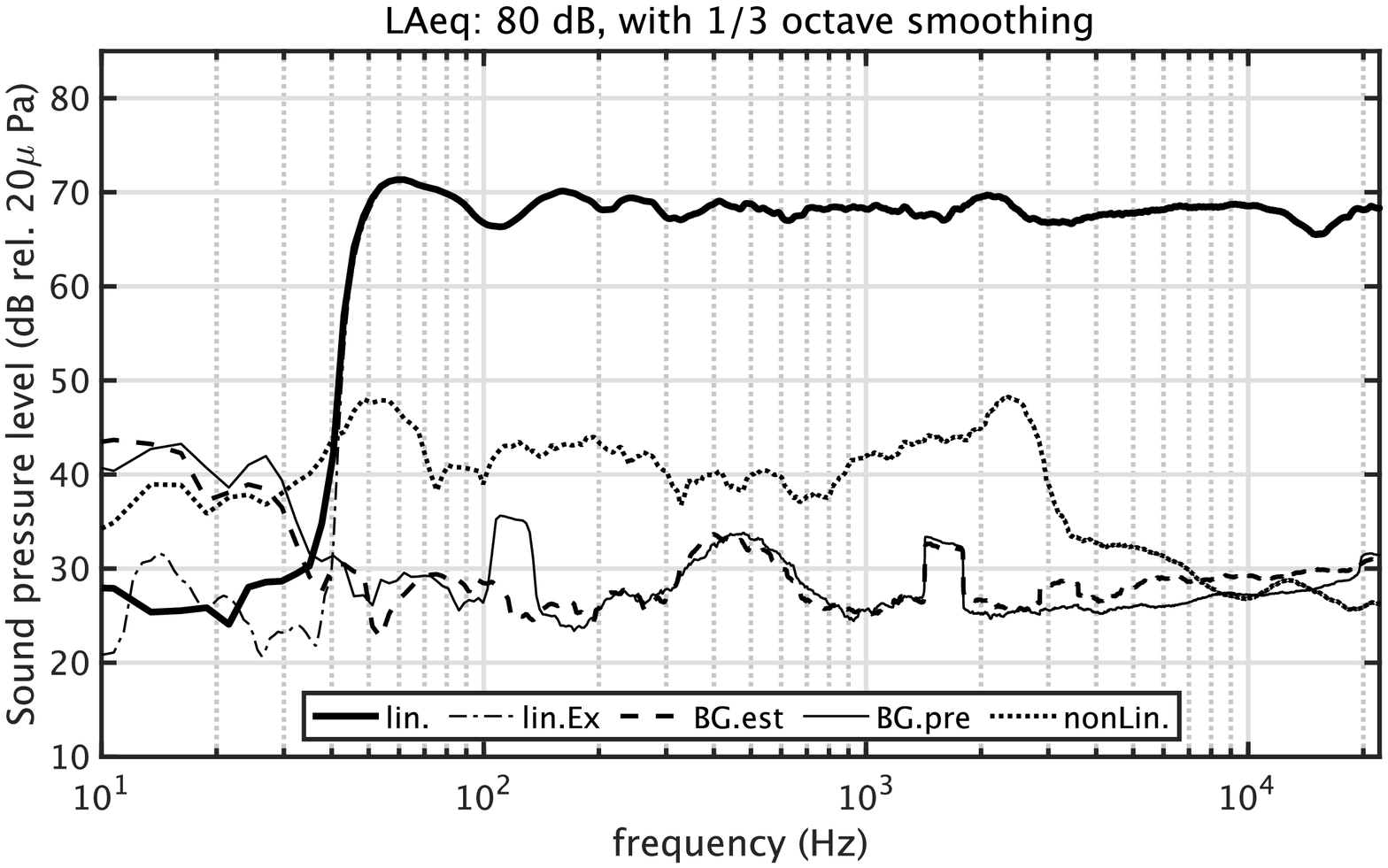}\\
\vspace{2mm}
\hfill\includegraphics[width=0.9\hsize]{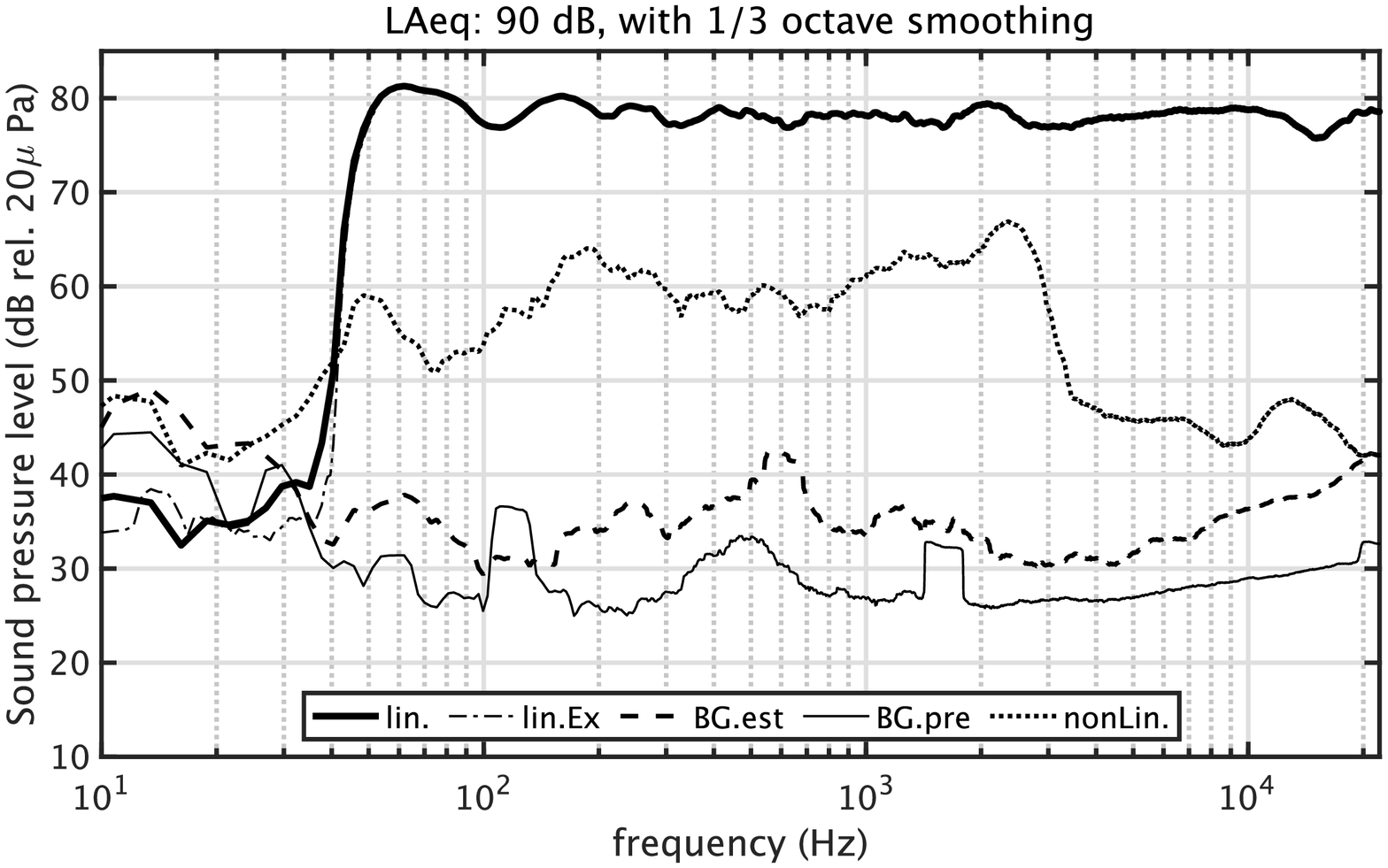}\\
\vspace{-2mm}
\caption{Frequency responses of a small powered loudspeaker system at three ($L_{Aeq}$: 70, 80, and 90~dB) sound pressure levels.
In a living room located in residential area.}
\label{fig:orthFresp}
\end{center}
\vspace{-4mm}
\end{figure}
Figure~\ref{fig:orthFresp} shows the measured results . 
We measured responses using three sound pressure levels (70~dB, 80~dB, and 90~dB in $L_{Aeq}$) at the microphone position.

The results for 70~dB indicates that the acoustic system operating at this sound pressure is practically indistinguishable from a linear time-invariant system because the background noise level masks the contribution from nonlinearity.
The results for 80~dB shows the contribution of nonlinearity clearly.
The spectrum of ``nonlinear'' component stands out from the background noise.
The results for 90~dB shows stronger nonlinearity contribution and increase of the estimated random component from the background noise measured during the preceding quiet region.
Note that the background noise in the region is same to the noise for 80~dB case, suggesting extra component exists in 90~dB case.
This increase in noise may indicate that the loudspeaker is generating extra random sounds caused by strong drive of moving parts.
Random component deviation in the high ($> 3$~kHz) may be the result of phase modulation caused by sound propagation time modulation due to air flow\cite{guidorzi2015impulse,kawahara2019apsipa}.
We tested several loudspeaker systems and found this random component increase in high frequency end exists in all high sound pressure level cases.
We also found that this increase is small with reduced air flow conditions.
We will elaborating systematic tests on this issue and will report elsewhere.

\begin{figure}[tbp]
\begin{center}
\includegraphics[width=0.99\hsize]{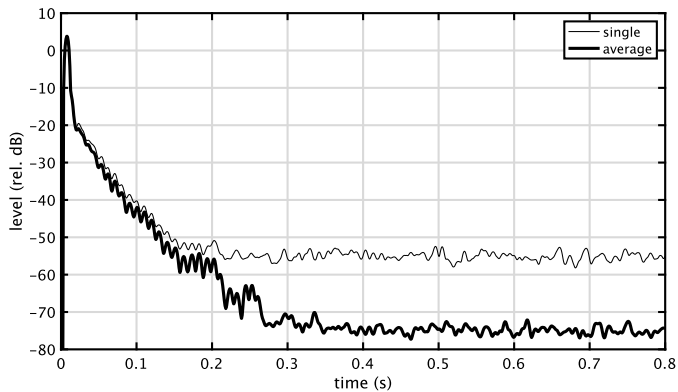}\\
\vspace{-2mm}
\caption{Power response example in a living room.}
\label{fig:livingRoomDecaiSample}
\end{center}
\vspace{-5mm}
\end{figure}
Figure~\ref{fig:livingRoomDecaiSample} shows the single and averaged power response in a Japanese living room in a residential area.
The sound pressure level in A-weight at 20~cm from the loudspeaker was 80~dB.
The responses are expanded versions.
The allocation interval was 0.2~s, and the expanded response has 0.8~s.
The thick line shows the average result of 100 measurements using different random number seeds.
In the plateau region, the average response level is 20~dB lower than that of a single measurement.
This 20~dB suppression experimentally supports that the high degree of variability of each FVN set made nonlinearity effect is independent of each other.

\section{Realtime and interactive tools}\label{ss:realtimetool}
We implemented an interactive and realtime tool for acoustic measurement using FVN-based methods described in the previous sections.
\begin{figure}[tbp]
\begin{center}
\includegraphics[width=0.99\hsize]{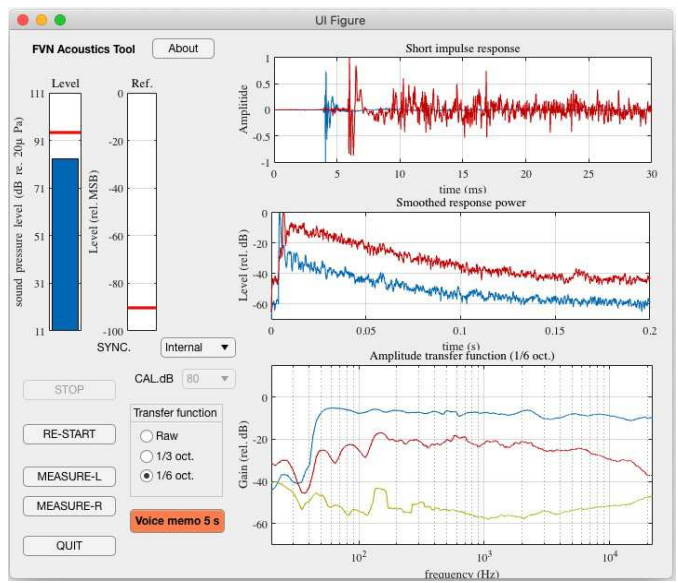}\\
\vspace{-2mm}
\caption{GUI snapshot of the interactive and realtime acoustic measurement using FVN-based procedures.}
\label{fig:GUISnapShot}
\end{center}
\vspace{-5mm}
\end{figure}
Figure~\ref{fig:GUISnapShot} shows a snapshot of the GUI of the tool.
We implemented the tool with MATLAB and open-sourced.
We also compiled it to prepare a standalone application for macOS and Windows10.
The tool and applications with a quick-start manual are accessible on our GitHub repository.

This section introduces this tool briefly.
Please refer to the documents in the repository.
Two bar graphs in the top left corner are the input level monitors of the left and the right channels.
While the start-up process user can calibrate the tool's sensitivity and make the left plot indicate the calibrated sound pressure level.
The default mode of the tools is a realtime interactive measurement of impulse responses from two speakers.
It implements the measurement using two FVN sequences described in~\ref{ss:twoFVNseq}.

Three panels on the right side show the measurement results in realtime.
The top panel shows the initial 30~ms of the impulse responses of the left and the right channels.
The middle panel shows the smoothed power of each impulse response.
The display's viewing length is equal to the repetition period of the unit FVNs ($n_\mathrm{o}/f_s$~s).
The bottom plot shows the smoothed power spectra of each response and the extra component calculated using the third unit FVN.
This third component represents deviations from linear time-invariant responses, which consist of background noise and interfering sounds, nonlinear component, and effects of movement of the loudspeakers and the microphone.

Buttons in the lower left side start offline measurements using four FVN sequences described in~\ref{ss:fvnMes4seq}.
The ``MEASURE-L'' and ``MEASURE-R'' button test the left and the right loudspeaker system, respectively.
The measurement started by ``MEASURE-L'' yields the analysis results shown in Fig.\ref{fig:orthFresp}.
The tool saves the visualization image, and the analysis results in files with unique names.

\section{Discussion}
The proposed method is general and applies to other fields.
The first author conducted auditory feedback experiments for measuring interactions between speech production and auditory perception\cite{kawahara1994interactions,Kawahara1996is}.
We used a pseudo-random noise (MLS) for modulating $f_\mathrm{o}$\footnote{We use $f_\mathrm{o}$ for representing the fundamental frequency 
instead of using F0. Please refer to the discussions in JASA forum article\cite{titze2015jasaforum}.} 
of the fed-back speech to make perturbation patter not to be predictable.
This auditory feedback control of voice $f_\mathrm{o}$ consists of nonlinear biological systems.
Application of the proposed method for auditory feedback research\cite{MURBE200244} enables us full control of experimental details and enables the decomposition of the linear time-invariant response and other nonlinear and time-varying responses.

An enormous amount of speech samples are available online today.
However, their recording conditions usually do not satisfy the recommendations for voice research\cite{svec2010ajsp,Rita2018ajsp}.
Moreover, they are usually using lossy audio digital compression\cite{moriya2016progress}.
The proposed method is useful for assessing degradations introduced by these non-ideal recording and coding conditions by decomposing the response into the linear time-invariant responses and other nonlinear and time-varying responses.
This application is useful for selecting usable materials from existing online speech resources for voice research.

The FVN-based method is new, and there are many issues.
For example, the six-term cosine series provided a feasible solution for designing the unit FVN.
However, it does not assure that the function is optimum.
Optimum windows proposed for phase-based analyses\cite{kusano2020maximally} for generalizing for higher-order continuities is a promising direction.
There is another possibility that the optimal energy concentrated bounded function\cite{slepian1961prolate} is also optimum for FVN design.
These are topics of further research.

\section{Conclusions}
We introduced a new acoustic measurement method that can simultaneously measure system attributes. They are linear time-invariant, and the other nonlinear and time-varying attributes. The method uses a set of orthogonal sequences made from a set of unit FVNs, a new member of the TSP. FVN has a unique feature that other TSP members do not. A high degree of design freedom makes the simultaneous measurement of acoustic attributes possible without introducing extra equipment. We introduced two useful cases using two and four orthogonal sequences and illustrates their use using simulations and acoustic measurement examples. We developed an interactive and realtime acoustic analysis tool based on the proposed method. We made it available in an open-source repository\cite{kawahara2020git}. The proposed response analysis method is general and applies to other fields, such as auditory-feedback research and assessment of sound recording and coding.

\section*{Acknowledgment}
This research is supported by Kakenhi 16H01734, 18K00147, 19K21618 of JSPS.
We appreciate Yutaka Kaneda for in-depth discussions on TSP-based acoustic measurements.

\bibliographystyle{IEEEtran}

\bibliography{APSIPA20FVNext.bib}

\end{document}